\documentclass[%
 reprint,
 amsmath,amssymb,
 aps,
]{revtex4-2}

\usepackage{graphicx}
\usepackage{subcaption}
\usepackage{dcolumn}
\usepackage{bm}
\usepackage{xcolor}
\usepackage{bbold}
\usepackage[utf8]{inputenc}
\usepackage{amsmath}

\newcommand{\cket}[1]{\left|{#1}\right)}
\newcommand{\cbra}[1]{\left({#1}\right|}

\begin{document}

\preprint{APS/123-QED}

\title{Improving quantum walk metrology with split-step quantum walk}

\author{Majid Moradi}%
 \email{majid.moradi@shahroodut.ac.ir}
 \author{Mostafa Annabestani}
 \email{annabestani@shahroodut.ac.ir}
\affiliation{%
 Faculty of Physics, Shahrood University of Technology\\
  Shahrood, Iran
}%

\date{\today}

\begin{abstract}
A new estimation scheme based on the split-step quantum walk (SSQW) revealed that by just setting a single parameter, SSQW can potentially achieve quantum Crame\'r-Rao bound in multiparameter estimation. This parameter even does not involve the parameterization but the initial state and unlike ordinary Quantum walk (OQW) there is no necessity for an entangled initial states or even a parameter dependent initial state. The rigorous analytic equations derived in this study revealed that SSQW surpasses OQW in achievable precision of multiparameter estimation in almost all possible scenarios. Furthermore, in single parameter estimation, the extra parameter can be used to tune the dynamics of the walk in such a way to enhance the precision of the estimation through maximizing the elements of quantum Fisher information matrix. The results of this study indicate that SSQW can remarkably improve the estimation schemes through its rich topological properties.
	
\end{abstract}

\maketitle


\section{\label{sec:level1}Introduction}
Many parameters in physics and quantum mechanics cannot be accessed directly and therefore it is not possible to measure such parameters without intermediaries \cite{Giovannetti2006, Taylor2016}, so estimation and metrology can be the only choice. Metrology is of high importance in science since it identifies the most precision which can be achieved in an estimation process \cite{Helstrom1968}.
Helstrom made one of the first contributions to quantum estimation and derived quantum Crame\'r-Rao bound \cite{Helstrom1968}. Also Personick studied the application of quantum estimation in performing analog communication through quantum channels \cite{Personick1971}. Since then, different methods in quantum estimation have been studied \cite{YOUNG197525,Matsumoto1997,Giovannetti2006,Paris2009} and now it has many applications in science. Some of the highlights are in biology \cite{Taylor2016}, LIGO (The Laser Interferometer Gravitational-Wave Observatory) \cite{Caves1981,Kimble2001}.

Discrete-time quantum walk (DTQW) as quantum version of classical random walk, where the particle hops between sites was first introduced by early works of \cite{Aharonov1993,aharonov2001,Ambainis2001,WATROUS2001,Kempe2003}. DTQW has been studied widely ever since regarding different aspects extensively: DTQW on line \cite{Ambainis2001,Kempe2003}, 2D \cite{Mackay2002,Carneiro2005,Oliveira2006,Watabe2008,Schreiber2012,Annabestani2010} and 3D \cite{Ehrhardt2020} lattices and hypercubes \cite{Moore2002,Marquezino2008}. People also considered entanglement in quantum walks \cite{Omar2006} different topologies including quantum walk on cycles \cite{BEDNARSKA2003,DUKES2014}, with boundaries \cite{BACH2004,Kwek2011,Kuklinski2020} and M\"obius quantum walk \cite{Moradi2017}.

Quantum walks also have been used as a computation tool and algorithm \cite{Feynman1986,Shenvi2003,Childs2004,Zhou2021}. One of the implications in realization of a quantum computer is the abiliy for imittating quantum systems \cite{Mallick2015}. Kitagawa studied different types of 1D quantum walks \cite{Karski2009,Zahringer2010,Schmitz2009,Schreiber2010,Broome2010,Ryan2005} and showed that quantum walks can be used to explore topological phases \cite{Kitagawa2010}. Arrighi et.al. showed that Dirac equation can be modeled as a discrete quantum walk \cite{Arrighi2014}. Mallick and Chandrashekar used split-step quantum walk to recover Dirac cellular automata \cite{Mallick2015}. Quantum walk also found its way in metrology and estimation to investigate quantum systems. Considering that the Hilbert space of discrete time quantum walk consists of two distinct subspaces, coin and position and after a few first steps these subspaces get entangled with each other, Singh et. al. come up with this novel idea that the parameters in coin can be estimated by performing measurement on positions space \cite{Singh2019}. Although most of the literature in quantum estimation theory are focused on single parameter estimation\cite{Helstrom1969, Braunstein1994, Toth2014}, but there are an increasing number of studies regarding multiparameter quantum estimation \cite{Berry2015, Baumgratz2016, Crowley2014}. Among them, Annabestani et.al. generalized the idea of using quantum walk in quantum metrology to multiparameter quantum metrology \cite{Annabestani2022} and as a case study they used quantum walk for joint estimation of mass and charge in Dirac equation.

\section{\label{QauntumMetrology}Quantum Metrology}

There are quantities for them it is impossible to make direct measurement. Also for some quantum mechanics quantities, there is no corresponding quantum observable \cite{Paris2009}, making quantum estimation a necessity. Quantum estimation is the process of finding the best estimation for a parameter in density operator \cite{Helstrom1969}. In other words, quantum estimation is finding the value for a parameter indirectly by means of making measurement on a quantum system affected by that parameter. 

A quantum metrologicl process involves four stages:
\begin{itemize}
	\item Preparation of the state. This state is going to be used as a probe.
	\item Parametrization. The parameter is going to be encoded on the probe by means of a parameter-dependant dynamics such as evolution under the action of a Hamiltonian or Liouvillian \cite{Annabestani2010}.
	\item Measurement. It is also important because getting close to Crame\'r-Rao bound as much as possible is dependant on performing a measurement in optimal bases \cite{Humphreys2013}.
	\item Classical estimation \cite{Liu2020}.
\end{itemize}

Suppose it is desirable to estimate the unknown parameter $\theta$ through a known conditionl density operator $\rho_{\theta}$ which depends on $\theta$. Holestrom found an inequality similar to Crame\'r-Rao bound \cite{Cramer1946} in quantum estimation \cite{Helstrom1968}, which limits the ultimate achievable precision of the unbiased estimator $\hat{\theta}$:
\begin{equation} \label{CramerRao}
	\sigma^2 \left( \hat{\theta} \right) \geq \mathcal{F} \left( \theta \right)^{-1} 
\end{equation}
where $\sigma^2 \left( \hat{\theta} \right)$ is the variance of the estimator and $\mathcal{F}$ is the so-called quantum Fisher information \cite{Braunstein1994} defined as
\begin{equation}
	\mathcal{F} \left( \theta \right) = Tr \left( \rho_{\theta} L_{\theta} ^2 \right).
\end{equation}
$L_{\theta}$ known as symmetric logarithmic derivative (SLD), is defined as
\begin{equation}
	\frac{\partial \rho_{\theta}}{\partial \theta} = \frac{1}{2} \left\{ L_{\theta} , \rho_{\theta} \right\} = \frac{1}{2} \left( L_{\theta} \rho_{\theta}+ \rho_{\theta} L_{\theta} \right)
\end{equation}
It shows that for optimal measurement the projector operator should be made over eigenstates of $L_{\theta}$, but $L_{\theta}$ may not always correspond to an optimal observable \cite{Paris2009}. It should be noticed that optimal measurement saturates Crame\'r-Rao inequality Eq. \eqref{CramerRao}.

Fortunately this formalism can be generalized to multiparameter quantum estimation \cite{Holevo1973,Paris2009,Humphreys2013,Szczykulska2016}. In mulltiprameter quantum estimation, multiple parameters are going to be encoded on the probe state by means of action of the quantum dynamics and the estimation is going to be performed simultaneously on multiple parameters \cite{Ragy2016}.

Suppose we have the density matrix $\rho_{\Theta}$ as probe and $\Theta = \left(\theta_{1},\theta_{2}, ..., \theta_{n}\right) , \theta_{i} \in \mathbb{R}$ is the set of parameters we are going to simultaneously estimate. The Cra\'mer-Rao bound for this multiparameter estimation is a generalization of Cra\'mer-Rao inequality for single parameter estimation \cite{Hayashi2008}, i.e.
\begin{equation}\label{MultiCramerRao}
	Cov \left( \Theta \right) \geq \mathcal{F}^{-1}
\end{equation}
where $ Cov \left( \Theta \right)$ indicates the covariance matrix of the size of number of parameters under estimation and $\mathcal{F}$ is quantum Fisher information matrix (QFIm) defined as
\begin{equation} \label{QFIm}
	\mathcal{F}_{\mu \nu} = \frac{1}{2} Tr \left( \rho_{\Theta} \left\{ L_{\mu} , L_{\nu} \right\} \right)
\end{equation}
where $L_{\mu}$ is the SLD with respect to parameter $\mu$.

Because the operators generally, do not commute, quantum Crame\'r-Rao bound for multiple quantum estimation in Eq. \eqref{MultiCramerRao} cannot be achieved always. It means, it is not generally possible to measure two observable with arbitrary precision, causing incompatibility of optimal measurement for distinct parameters \cite{Ragy2016}. Given any cost matrix $G$, the precision in estimation will be bounded by
\begin{equation}
	Tr \left( G Cov \left( \Theta \right) \right) \geq Tr \left( G \mathcal{F}^{-1} \left( \Theta \right) \right) = C^{S} \left( \Theta , G \right)
\end{equation}
where $C^{S} \left( \Theta , G \right)$ is known as symmetric bound \cite{Ragy2016,Annabestani2022}. Holevo calculated a more powerful bound $C^{H} \left( \Theta , G \right)$. Holevo Cr\'amer-Rao bound has been upper bounded by \cite{Carollo2019}
\begin{equation}
	C^{S} \left( \Theta , G \right) \leq C^{H} \left( \Theta , G \right) \leq \left( 1 + \mathcal{R} \right) C^{S} \left( \Theta , G \right)
\end{equation}
where $\mathcal{R}$ is the incompatibility between parameters $\mu$ and $\nu$
\begin{equation}\label{incompatiblity}
	\mathcal{R} = \|i\mathcal{F}^{-1} \mathcal{D} \|_{\infty}
\end{equation}
where $\|.\|_{\infty}$ denotes the largest eigenvalues of a matrix \cite{Carollo2019} and $\mathcal{D}$ is the mean Uhlmann curvature (MUC) matrix
\begin{equation} \label{Uhlmann}
	\mathcal{D}_{\mu \nu} = -\frac{i}{2} Tr \left( \rho_{\Theta} \left[ L_{\mu}, L_{\nu} \right] \right).
\end{equation}
$\mathcal{R}$ quantifies the amount of incompatibility of two parameters $\mu$ and $\nu$ within a parameter estimation model, between $0$ for full compatibility and $1$ indicating full incompatibility \cite{Carollo2019, Annabestani2022,Belliardo2021}.

Generally  in quantum estimation, it is not possible to perform optimal measurements for individual parameters due to the fact the corresponding SLDs for those parameters do not commute in general. Mean Uhlmann curvature captures the geometry of such a fundamental limitation and a vanishing MUC indicates that Crame\'r-Rao bound is attainable \cite{Ragy2016, Carollo2018} and an asymptotically compatible multiparameter quantum estimation scheme. A diagonal Fisher information matrix may also be a strong indicator of compatibility, because zero off-diagonal elements shows that the measurement can be performed simultaneously on both parameters with optimal precision determined by QCRB. But one should take care, because although diagonal QFIm is a strong signal for compatibility, it does not always guarantee a vanishing MUC and a compatible quantum multiparameter estimation. Thereupon, a quantum multiparameter estimation is compatible if and only if MUC vanishes \cite{Carollo2018}. For such a scenario, Holevo QCRB coincides with the standard QCRB \cite{Li2022} and it is possible to achieve the same precision as of individual parameter estimation. Based on this argument, an optimal estimation scheme is achievable if
\begin{itemize}
	\item [1)] The MUC vanishes. Holevo QCRB and standard QCRB meet.
	\item [2)]	QFIm to be maximized. A larger QFIm results in tighter QCRB, see (\ref{MultiCramerRao}).
	\item [3)]	There should be a single probe state which maximizes quantum Fisher information matrix (QFIm) for all parameters simultaneously \cite{Annabestani2022}.
\end{itemize}
Once all these conditions are satisfied the precision of multiparameter and separate estimation would be the same \cite{Ragy2016, Annabestani2022}. Such models are called compatible models \cite{Ragy2016}.

\section{\label{DTQW}Discrete-Time Quantum Walk}

1D classical random walk (CRW) is indeed a Bernoulli experiment followed by a 1D move in position space \cite{Pearson1905,Spitzer2001}. 1D discrete-time quantum walk (DTQW) is a generalization of 1D CRW, where the most remarkable different aspect of 1D DTQW is that the walker can move in a superposition of all possible directions \cite{Nayak2000}

The characteristic space of 1D DTQW is consisted of two Hilbert sub-spaces $\mathcal{H} = \mathcal{H}_{p} \otimes \mathcal{H}_{c}$, where $\mathcal{H}_p$ is the position space with the basis $\left\{ \left| x \right\rangle | x \in \mathbb{Z} \right\}$ and $\mathcal{H}_c$ denotes the coin space which decides on the walker's movement direction $\left\{ \left| j \right\rangle | j \in \left\{ 0,1 \right\} \right\}$, so the state of the walker can be denoted as
\begin{equation}
	\left| \Psi \left(0\right) \right\rangle = \sum_{x,j} \psi_{x,j} \left(0\right) \left| x \right\rangle_{p} \left| j \right\rangle_{c}.
\end{equation}
The evolution operator of walk
\begin{equation}
	U_\Theta = S \left( \mathbb{1} \otimes C_\Theta \right)
\end{equation}
is composed of two steps, firstly the coin operator $C_\Theta$ acts on the coin part $ \left| j \right\rangle_{c}$ creating a superposition of $\left| 0 \right\rangle$ and $\left| 1 \right\rangle$. After that a shift operator $S$ moves the walker based on its coin status.
The idea for using DTQW as a probe for quantum estimation is that the desired parameters for estimation can be encoded to the walker's state $\left| \Psi \right\rangle$ by means of the evolution operator $U_\Theta$ or more specifically the coin operator $C_\Theta$ in the evolution operator \cite{Singh2019}. 

Without loss of generality for 1D QW we can use a SU(2) operator for the most general form of coin operator as a
\begin{equation}
	C_\Theta =\left( \begin{array}{cc}
		e^{i \alpha} \cos(\theta) & e^{i \beta} \sin(\theta) \\
		-e^{-i \beta} \sin(\theta) & e^{-i \alpha} \cos(\theta)
	\end{array}\right)
\end{equation}
which has three parameters $\Theta=\left\{\alpha, \beta,\theta \right\}$. So it seems, 1D QW can be used to multi-parameter estimation up to three parameters, but it is shown that only two independent parameters can be estimated by 1D QW \cite{Annabestani2022}. The one parameter estimation ($\alpha=\beta=0$) by quantum walk has been studied before \cite{Paris2009}.

1D split-step quantum walk (SSQW) is another type of 1D QWs, where the walk is splitted by applying two different coins \cite{Kitagawa2010}
\begin{equation}
	C_{i}=\left( \begin{array}{cc}
		\cos(\frac{\theta_{i}}{2}) & -\sin(\frac{\theta_{i}}{2}) \\
		\sin(\frac{\theta_{i}}{2}) & \cos(\frac{\theta_{i}}{2})
	\end{array}\right) , i = 1 , 2
\end{equation}
and subsequently two different shift operators
\begin{align}
	S_1 &= \sum_{x} \left( I \otimes \left|1 \right\rangle \left\langle 1 \right| + \left|x+1 \right\rangle \left\langle x \right| \otimes \left|0 \right\rangle \left\langle 0 \right| \right) \\
	S_2 &= \sum_{x} \left( \left|x-1 \right\rangle \left\langle x \right| \otimes \left|1 \right\rangle \left\langle 1 \right| +I \otimes \left|0 \right\rangle \left\langle 0 \right| \right)
\end{align}
So, the evolution operator would be
\begin{equation}\label{U}
	U_{\theta_{1} , \theta_{2}} = S_{2} \left( I \otimes C_{2} \right) S_{1} \left( I \otimes C_{1} \right).
\end{equation}
The state of the walker after $t$ steps is
\begin{equation}\label{finaltateOfQW}
	\left| {{\Psi _\Theta }\left( t \right)} \right\rangle  = U_\Theta ^t\left| {\Psi \left( 0 \right)} \right\rangle. 
\end{equation}
Due to the valuable features of the split-steps quantum walk (SSQW) model, it is an interesting topic to be studied by researchers. Dirac quantum cellular automaton \cite{Mallick2016}, Simulating Dirac Hamiltonian in curved space-time \cite{Mallick2019} and Exploring topological phases \cite{Kitagawa2010} are just a few examples of papers based on SSQW.

In this paper we have use SSQW model for quantum meteorology to show how this model can optimize the estimation.

\section{\label{BasicFormalism} Basic Formalism}
Generally the final state of walker in Eq. \eqref{finaltateOfQW} contains embedded unknown parameters $\Theta$ and can be used for estimation, but the non-diagonal large matrix $U_\Theta$ is very hard to handle. Fortunately the Fourier transformation
\begin{equation}
	\left| x \right\rangle  = \int_{ - \pi }^\pi  {\frac{{dk}}{{2\pi }}{e^{ - ikx}}\left| k \right\rangle }.
\end{equation}
can convert Eq. \eqref{U} into block-diagonal form
\begin{equation}
	{U_\Theta } = \int_{ - \pi }^\pi  {\frac{{dk}}{{2\pi }}\left| k \right\rangle \left\langle k \right| \otimes {u_k}\left( \Theta  \right)},
\end{equation}
where
\begin{equation}\label{uk}
	{u_k}\left( \Theta  \right) = \left( {\begin{array}{*{20}{c}}
			1&0\\
			0&{{e^{ik}}}
	\end{array}} \right){C_2 } + \left( {\begin{array}{*{20}{c}}
			{{e^{ - ik}}}&0\\
			0&1
	\end{array}} \right){C_1 }.
\end{equation}
So the final state of the walker after \textit{t} steps is
\begin{equation} \label{PsiTermsofPhi}
	\left| {{\Psi _\Theta }\left( t \right)} \right\rangle  = \int_{ - \pi }^\pi  {\frac{{dk}}{{2\pi }}\left| k \right\rangle  \otimes 	{u_{k}^{t}} \left| {\varphi _k\left( 0  \right)} \right\rangle }
\end{equation}
where $\left| {{\varphi _k}\left( 0 \right)} \right\rangle =\left\langle {k}
\mathrel{\left | {\vphantom {k {\Psi \left( 0 \right)}}}
	\right. \kern-\nulldelimiterspace}
{{\Psi \left( 0 \right)}} \right\rangle$ and ${u_{k}}\equiv{u_{k}}\left( \Theta \right)$ for simplicity.

Recently M. Annabestani et al. \cite{Annabestani2022} show that for unitary evolution of quantum walk, the Fisher information and mean Uhlmann curvature matrices can be written as

 \begin{widetext}
	\begin{align} \label{FisherUhlmannBasic}
		\mathcal{F}_{\mu \nu} &= 4 \mathcal{R} \left(\int_{-\pi}^{\pi} \frac{dk}{2 \pi} \left\langle A_{k}^{'} \left( O_{\mu} ^{\dag}  \right) A_{k}^{'} \left( O_{\nu}  \right) \right\rangle_{t} - \left( \int_{-\pi}^{\pi} \frac{dk}{2 \pi} \left\langle A_{k}^{'} \left( O_{\mu} ^{\dag}  \right) \right\rangle_{t} \right) \left( \int_{-\pi}^{\pi} \frac{dk}{2 \pi} \left\langle A_{k}^{'} \left( O_{\nu}  \right) \right\rangle_{t} \right) \right) \nonumber \\
		\mathcal{D}_{\mu \nu} &= 4 \mathcal{I} \left( \int_{-\pi}^{\pi} \frac{dk}{2 \pi} \left\langle A_{k}^{'} \left( O_{\mu} ^{\dag}  \right) A_{k}^{'} \left( O_{\nu}  \right) \right\rangle_{t} - \left( \int_{-\pi}^{\pi} \frac{dk}{2 \pi} \left\langle A_{k}^{'} \left( O_{\mu} ^{\dag}  \right) \right\rangle_{t} \right) \left( \int_{-\pi}^{\pi} \frac{dk}{2 \pi} \left\langle A_{k}^{'} \left( O_{\nu}  \right) \right\rangle_{t} \right) \right),
	\end{align}
\end{widetext}
where $O_{\mu} = u_{k}^{\dagger} \partial_{\mu} u_{k}$, super operator of $A_k\left(O\right)\equiv u_{k} O u_{k}^{\dagger}$ and
 \begin{equation}
 	 A_{k}' \left( O_{\mu} \right)=\sum_{m=0}^{t-1} A_{k}^{m+1} \left( O_{\mu} \right)\equiv\sum_{m=0}^{t-1} u_{k}^{m+1} O_{\mu} {u_{k}^{m+1}}^{\dagger}. 
 \end{equation}

 They have also shown that these matrices converge quickly to the asymptotic form of  
 
 \begin{widetext}
 	\begin{align}\label{assymptotic_form}
 		\mathcal{F}_{\mu \nu} &= t^{2} \left( \int_{-\pi}^{\pi} \frac{dk}{2 \pi} \left( O_{\mu} \right| {\tilde{A}}_{k}^{\mathbb{1}} \left| O_{\nu} \right) - \left( \int_{-\pi}^{\pi} \frac{dk}{2 \pi} \left( O_{\mu} \right| {\tilde{A}}_{k}^{\mathbb{1}} \left| \rho_{0} \right) \right) \left( \int_{-\pi}^{\pi} \frac{dk}{2 \pi} \left( \rho_{0} \right| {\tilde{A}}_{k}^{\mathbb{1}} \left| O_{\nu} \right) \right) \right) + O \left( t \right) \nonumber \\
 		\mathcal{D}_{\mu \nu} &= \int_{-\pi}^{\pi} \frac{dk}{2 \pi} \left( \vec{ o_{\mu}^{'}} \times \vec{ o_{\mu}^{'} } \right) . \vec{r}
 	\end{align} 
 \end{widetext} 
which utilizes the vectorization as
\begin{equation}\label{4vector-O}
	\left| O \right) = \begin{pmatrix}
		o_0 \\ o_1 \\ o_2 \\ o_3
	\end{pmatrix} \equiv \left( o_0 , \vec{o} \right)^{T}
\end{equation}
where $\vec{o}$ is the Bloch vector, defined as 
\begin{equation}\label{BlochRep}
	O = \frac{1}{2} \sum_{i=0} \left( o_{i} \sigma_{i} \right)
\end{equation}
in which $o_i = Tr \left( O \sigma_i \right)$, $\sigma_i$s are Pauli matrices and $\sigma_0 = \mathbb{1}$. 

${\tilde{A}}_{k}^{\mathbb{1}}$ in Eq. \eqref{assymptotic_form} is the sub-space of $\tilde{A_{k}}$ in which it's eigenvalues are equal to one. Note that {\textasciitilde} has been used to indicate  the vectorized form  of super operators (${A}_k\left(O\right)\equiv\tilde{A}_k\left|O\right)$).

We used this formalism to investigate the split-steps quantum walk as a topological quantum walk model \cite{Kitagawa2010} for estimation. Our investigations revealed that SSQW can significantly optimize the estimation process in different aspects. 
\section{\label{Results} Main Results}
Using Eq. \eqref{uk} leads to

\begin{equation}
u_k=\left(\begin{array}{cc}
	c_1 c_2 e^{-ik}-s_1 s_2  & -s_1 c_2 e^{-ik}  -c_1 s_2   
	\\
	 c_1 s_2 +e^{ik} s_1 c_2   & -s_1 s_2  +e^{ik} c_1 c_2   
\end{array}\right)
\end{equation}

where we used $c_i\equiv \cos(\frac{\theta_{i}}{2})$ and $s_i\equiv \sin(\frac{\theta_{i}}{2})$ for simplicity. Using $O_{\mu} = u_{k}^{\dagger} \partial_{\mu} u_{k}$ , Eq. \eqref{4vector-O} and Eq. \eqref{BlochRep}

\begin{equation}\label{4vector_O}
	\left|O_1\right)=-i\left(\begin{array}{c}
		0 
		\\
		0 
		\\
		1
		\\
		0 
	\end{array}\right), \left|O_2\right)=-i\left(\begin{array}{c}
	0 
	\\
	\left(2c_1^2-1\right)\sin\left(k\right)
	\\
	\cos\left(k\right)
	\\
	2c_1 s_1 \sin\left(k\right)
\end{array}\right),
\end{equation}

and by some calculations and simplifications one can reaches (see the explicit form of super-operator $\tilde{A}_{k}$ in Appendix.A )

\begin{widetext}
	\begin{align} \label{A1}
		\tilde{A}_{k}^{\mathbb{1}} = \frac{1}{N}\left(\begin{array}{cccc}
		N &0&0&0
		\\
		0&-s_1^{2} c_2^{2} \sin^{2}\left(k \right)  & s_1 c_2 \sin \left(k \right)  \left(c_1 s_2 +s_1 c_2 \cos \left(k \right)\right) &s_1 c_1 c_2^{2} \sin^{2}\left(k \right)    
		\\
		0&s_1 c_2 \sin \left(k \right)  \left(c_1 s_2 +s_1 c_2 \cos \left(k \right)\right) & -\left(c_1 s_2 +s_1 c_2 \cos \left(k \right)\right)^{2} & -c_1 c_2 \sin \left(k \right) \left(c_1 s_2 +s_1 c_2 \cos \left(k \right)\right)   
		\\
		0&s_1 c_1 c_2^{2} \sin^{2}\left(k \right)   & -c_1 c_2 \sin \left(k \right) \left(c_1 s_2 +s_1 c_2 \cos \left(k \right)\right)   & -c_1^{2} c_2^{2} \sin^{2}\left(k \right)  
	\end{array}\right)
\end{align}
\end{widetext}

where $N = \sin^2\left(\omega\right)$ and $\cos\left(\omega \right) = 
c_1 c_2 \cos\left(k \right)-s_1 s_2$.

To calculate Fisher information and Uhlmann curvature for SSQW (Eq. \eqref{assymptotic_form}), we need to calculate integrals of the form

\begin{equation}
	\int_{-\pi}^{\pi} \frac{dk}{2 \pi} \left(\bullet \right| {\tilde{A}}_{k}^{\mathbb{1}} \left| \bullet \right).
\end{equation}

Due to the presence of $N$ in denominator of ${\tilde{A}}_{k}^{\mathbb{1}}$, the integrand is not analytic for certain values. By Cauchy's residue theorem and defining $z=e^{ik}$, these integrals map to ones on a unit circle in $z$ plane having 4 simple poles

\begin{align}
	z_1&=\frac{1+s_1 s_2+\left|{s_1+s_2}\right|}{c_1 c_2} \nonumber \\
	z_2&=\frac{1+s_1 s_2-\left|{s_1+s_2}\right|}{c_1 c_2} \nonumber \\
	z_3&=\frac{-1+s_1 s_2+\left|{s_1-s_2}\right|}{c_1 c_2} \nonumber \\
	z_4&=\frac{-1+s_1 s_2-\left|{s_1-s_2}\right|}{c_1 c_2}.
\end{align}

Considering the sign of $\left(s_1-s_2\right)$, only two of the poles can be inside the unit circle. Note that, since $0\le \theta_i \le 2\pi$, $\left(s_1+s_2\right)$ is always positive.

\begin{figure}[!t]
	\includegraphics[width=0.6 \linewidth]{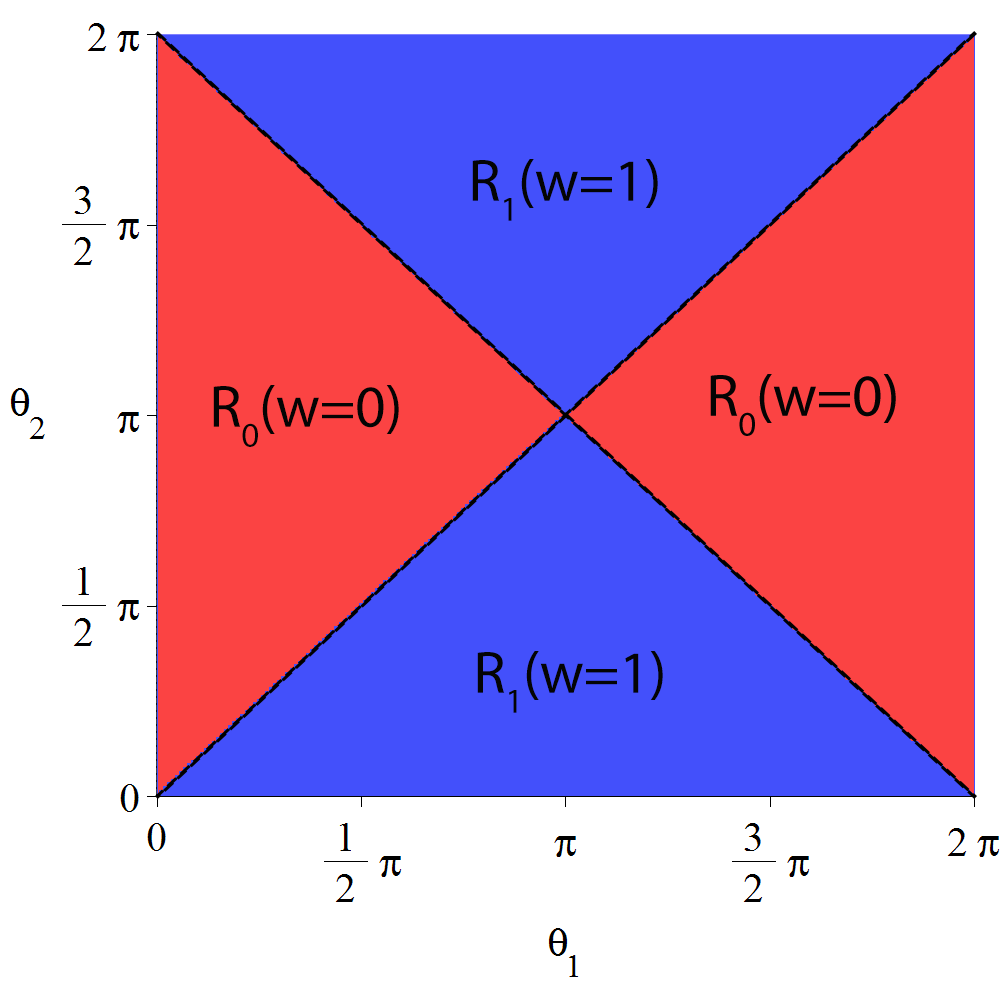}
	\caption{Blue region: $\left(s_1-s_2\right)>0$ in which  w=0 and Red region: $\left(s_1-s_2\right)<0$ in which  w=1}\label{zones}
\end{figure}

By these considerations we have two different regions: $R_0$ for $\left(s_1-s_2\right)< 0$ (red region) and $R_1$ for $\left(s_1-s_2\right)>0$ (blue region) (see Fig. \ref{zones}). Defining the initial state as $\cket{\rho_{0}} = \left(1, r_{1} , r_{2} , r_{3} \right)^{T}$, using Eq. \eqref{4vector_O} and Eq. \eqref{assymptotic_form}, we have 
\begin{widetext}
	\begin{equation}\label{F01}
		\begin{aligned} 
			\mathcal{F}^{0} = \left(\begin{array}{cc}
				\frac{s_2 - s_1^2}{c_1^2} &\frac{s_1 c_2}{c_1\left(s_2 + 1\right)}\\
				\frac{s_1 c_2}{c_1\left(s_2 + 1\right)}&\frac{s_2}{s_2+1}			
			\end{array}\right)-
		r_2^2\left(\begin{array}{cc}
			\left(\frac{s_2 - s_1^2}{c_1^2}\right)^2 &\frac{s_1 c_2}{c_1\left(s_2 + 1\right)}\left(\frac{s_2 - s_1^2}{c_1^2}\right)\\
			\frac{s_1 c_2}{c_1\left(s_2 + 1\right)}\left(\frac{s_2 - s_1^2}{c_1^2}\right)&\left(\frac{s_1 c_2}{c_1\left(s_2 + 1\right)}\right)^2			
		\end{array}\right)\\
			\mathcal{F}^{1} = \left(\begin{array}{cc}
				\frac{s_1}{s_1+1} &\frac{s_2 c_1}{c_2\left(s_1 + 1\right)}\\
				\frac{s_2 c_1}{c_2\left(s_1 + 1\right)}&\frac{s_1 - s_2^2}{c_2^2}			
			\end{array}\right)-
			r_2^2\left(\begin{array}{cc}
				\left(\frac{s_1}{s_1+1}\right)^2 &\frac{s_2 c_1}{c_2\left(s_1 + 1\right)}\left(\frac{s_1}{s_1+1}\right)\\
				\frac{s_2 c_1}{c_2\left(s_1 + 1\right)}\left(\frac{s_1}{s_1+1}\right)&\left(\frac{s_2 c_1}{c_2\left(s_1 + 1\right)}\right)^2			
			\end{array}\right),
		\end{aligned}
	\end{equation}
\end{widetext}
for region $R_0$ and $R_1$ respectively.

A very important fact is that, from all the variables in initial state, Fisher matrices are just dependent on $r_{2}$ in form of $r_2^2$. In fact the coefficient of $r_2^2$ in both $\mathcal{F}^0$ and $\mathcal{F}^1$ are positive matrices with positive diagonal elements indicating the diagonal elements $\mathcal{F}_{11}$ and $\mathcal{F}_{22}$ are maximized for $r_2=0$. So while the  most precise estimation (maximum diagonal elements) for ordinary QW needs an entangled initial state\textcolor{red}{\underline{s}} or at least a parameter-dependent initial state \cite{Annabestani2022}, the  most precise estimation in SSQW can be achieved for a large family of parameter-independent local initial states ($r_2=0$)!
     
\begin{figure}
	\begin{subfigure}[b]{0.35 \textwidth}
		\includegraphics[width=\textwidth]{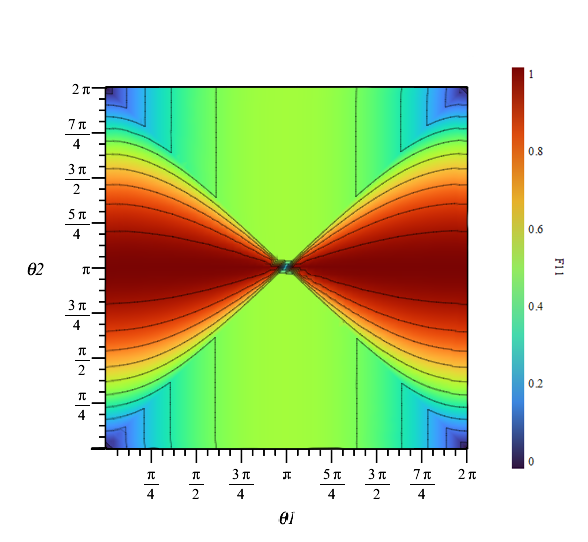}
	\end{subfigure}
	\begin{subfigure}[b]{0.3 \textwidth}
		\includegraphics[width= \textwidth]{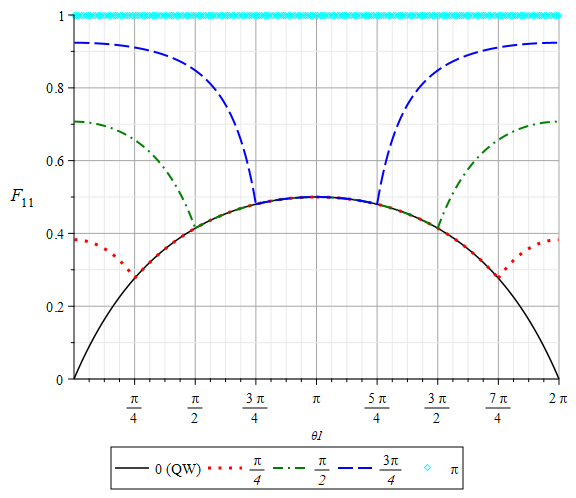}
	\end{subfigure}
	\caption{(Up) $\mathcal{F}_{11}$ versus $\left(\theta_{1},\theta_{2}\right)$ and (Down) cross section of $\mathcal{F}_{11}$ for various $\theta_{2}$. $\theta_{2}=0$ corresponds to ordinary QW. }\label{F11_3dand2d}
\end{figure}

Given quantum Fisher information's sensitivity to topological changes (the same is for the mean Uhlmann curvature) \cite{Yin2019} and also considering that SSQW exhibits strong topological features \cite{Kitagawa2010}, a discerning reader might ask whether there is any relationship between the behavior of QFIm of an estimation scheme based on SSQW and the topological properties of SSQW. From Fig.\ref{zones} it is obvious that QFIm elements can behave completely different in regions with different colors and these regions are exactly where SSQW exhibits specific topological properties characterized by a topological invariant known as winding number \cite{Kitagawa2010}. Apparently at border lines of these regions, system undergoes a topological phase transition. A deeper insight can be acquired by taking a look at Fig.\ref{F11_3dand2d}. It is clear that at border lines, Fisher information changes dramatically, pointing out a significant change in estimation properties. Indeed, these lines are characterized by $ \theta_{1} = \theta_{2} $ and $ \theta_{1} + \theta_{2} = 2 \pi $. One can see that by approaching these lines, the precision of SSQW-based estimation reduces and in worse case scenario, it equals to that of OQW-based estimation scheme Fig.\ref{F11_3dand2d}. So although topological invariants such as winding number do not appear explicitly in QFIm, but the results indicate much better multiparameter estimation results in sense of larger values of Fisher information for regions with non-trivial topological number denoted by $w=0$ and relatively weaker results for regions with trivial topological number i.e. $w=1$ (see Fig. \ref{F11_3dand2d}). Note that the worst case of estimation (minimum Fisher information) occurs at the critical points of the topological phase transition, which are exactly equal to those in OQW.

\subsection{Single-parameter estimation by SSQW}
In the context of parameter estimation, SSQW and ordinary QW can be used to estimate at most two parameters ($\theta_{1},\theta_{2}$) and ($\theta,\phi$) respectively.
Specifically, when estimating a single parameter, denoted as $\theta_{1}$, there exists a significant advantage in using SSQW over ordinary QWs.

This advantage stems from the ability to achieve greater Fisher information ($\mathcal{F}_{11}$) by tuning an additional parameter $\theta_{2}$ in SSQW (see Fig.\ref{F11_3dand2d}), while it is impossible for ordinary QW in which both of diagonal elements $\mathcal{F}_{\theta \theta}$ and $\mathcal{F}_{\phi \phi}$ just depend on $\theta$ and never change by tuning $\phi$ \cite{Annabestani2022}. 

More precisely, the ordinary quantum walk\textcolor{red}{\underline{s}} (OQW) is not optimal for estimating parameters with small values (the Fisher information will be small), whereas SSQW, by tuning $\theta_{2}$, can fix this problem perfectly (see Fig.\ref{F11_3dand2d}).

This optimization can be quantified by definition of average Fisher information 
\begin{eqnarray}
	\bar{\mathcal{F}_{x}}=\frac{1}{a-b}\int_{a}^{b}{\mathcal{F}_{xx} dx}
\end{eqnarray}
which determines the average precision of estimation for unknown parameter $x\in\left[a,b\right]$.

In the optimal strategy of estimation (maximum Fisher information) which is $r_2=0$ and $\vec{r}=\partial_{\theta}\hat{n}$ (see \cite{Annabestani2022}) in SSQW and QW respectively, using  Eq. \eqref{F01} leads to 
\begin{equation}
\bar{\mathcal{F}}^{SSQW}_{\theta_{1}}=1-\frac{2}{\pi}\left|\cos\left(\frac{\theta_{2}}{2}\right)\right|
\end{equation}
for SSQW, while 
\begin{equation}
	\bar{\mathcal{F}}^{QW}_{\theta}=1-\frac{2}{\pi}
\end{equation}
for ordinary QW (we used the explicit form of $\mathcal{F}$ from \cite{Annabestani2022}). It means that, when performing the same parameter estimation process with identical resources, SSQW (in average) can yield estimations up to approximately 2.75 times more precise compared to ordinary QW.

\subsection{Multi-parameter estimation by SSQW}

Obviously, both of SSQW and ordinary QW can be used for estimating at most two parameters. In this section we answer this question : which one is better?
 
In multi-parameter estimation, the precision of estimated parameters $\theta_{i}$, determined by $\mathcal{F}_i\equiv \mathcal{F}_{ii}$, has lower bound satisfying \cite{Guo2016.PhysRevA.93.042115}
\begin{equation}
	\mathcal{F}_i \mathcal{F}_j \ge \mathcal{D}_{ij}.
\end{equation}  
Since the inverse of $\mathcal{F}_i$ gives the lowest attainable bound for the variance of the estimated parameter (Eq. \ref{MultiCramerRao}), this inequality indicate\textcolor{red}{s} the upper bound for precision of estimation. So the higher $\mathcal{F}_i \mathcal{F}_j$, the more accurate is the estimation.
 
For the optimal strategy ($r_2=0$), it is clear from Fig. \ref{F11_3dand2d} and explicit form of $\mathcal{F}$ in Eq. \eqref{F01} that, the minimum value of $\mathcal{F}_{1}$ and $\mathcal{F}_{2}$ are on the line $\theta_{2}=\theta_{1}$ and $\theta_{2}=2\pi-\theta_{1}$ (which are interestingly, the critical points for topological phase transition of SSQW model \cite{Kitagawa2010}), so
 \begin{eqnarray}\label{OmegaSSQW}
 	\Omega_{min}^{SSQW}=\left.\mathcal{F}_1\mathcal{F}_2\right|_{\theta_2=\theta_1}=\left.\mathcal{F}_1\mathcal{F}_2\right|_{\theta_2=2\pi-\theta_1}=\left(\frac{s_1}{1+s_1}\right)^2
 \end{eqnarray}
On the other hand for optimal strategy in ordinary QW (see \cite{Annabestani2022}), we have 
\begin{eqnarray}\label{OmegaQW}
	\Omega^{QW}=\mathcal{F}_1\mathcal{F}_2=\frac{s_1\left(1-s_1\right)}{1+s_1}.
\end{eqnarray} 
The equation $\Omega_{min}^{SSQW}=\Omega^{QW}$ has two solutions $s_1=-g, s_1=-g^{-1}$ where $g=\frac{1+\sqrt{5}}{2}$ is the Golden ratio! (Note that the Golden ratio appears in OQW as a point in which both diagonal elements of Fisher information are maximum \cite{Annabestani2022}).

So for $\eta\le\theta_{2}\le\left(2\pi-\eta\right)$ always $\Omega_{min}^{SSQW}\ge \Omega^{QW}$, where $\sin\left(\frac{\eta}{2}\right)=-g^{-1}$.

\begin{figure}
	\includegraphics[width=0.8 \linewidth]{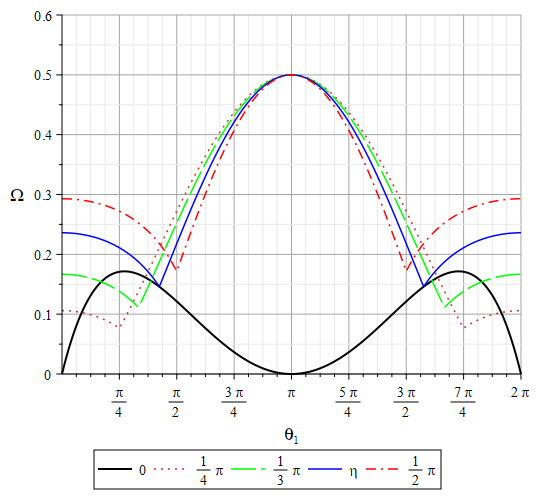}
	\caption{Cross section of $\Omega$ for various $\theta_{2}$. $\theta_{2}=0$ corresponds to ordinary QW.} \label{whichOneIsBetter2D}
\end{figure}

In summary, for estimation of two parameters, SSQW model is a better choice for almost all range of parameters. Only for small value of $\theta_{2}$  when $0 <\theta_1 <\eta$ or $2\pi-\eta < \theta_1 < 2\pi$ the ordinary QW has better performance a bit (see Fig. \ref{whichOneIsBetter2D} ).

\begin{figure}
	\includegraphics[width=0.8 \linewidth]{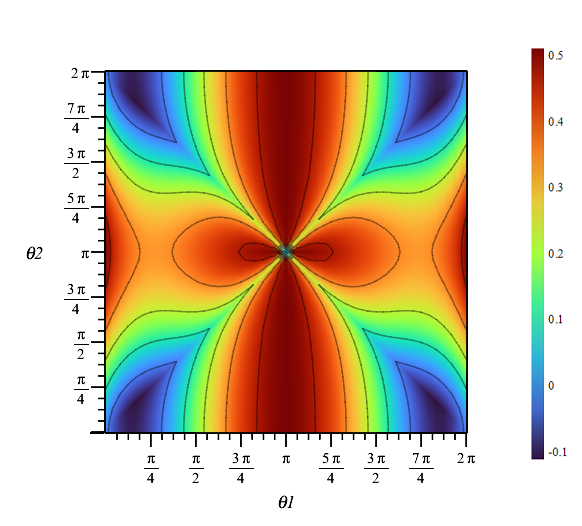}
	\caption{$\Omega^{SSQW}-\Omega^{QW}$ versus $\theta_{1}$ and $\theta_{2}$ }\label{whichOneIsBetter}
\end{figure}

To visualize the performance of the two quantum walk models, we examine the quantity ($\Omega^{SSQW}-\Omega^{QW}$), which has been plotted in Fig. \ref{whichOneIsBetter}. 

As we see, only within small regions (indicated by the dark blue color), this quantity is negative (reaching as low as -0.1). This implies that the ordinary quantum walk (QW) exhibits slightly better performance in these specific regions. It’s important to note that this area corresponds to the cases where the evolution of quantum walk closely approaches the identity. Consequently, the walk does not significantly deviate from its initial state.

\section{\label{Conclusion} Conclusion}
This study explored the landscape of application of split step quantum walk (SSQW) in quantum metrology and revealed the great potential of SSQW in optimization of estimation. These optimizations can be summarized in three general axes: First, while achieving the quantum Crame\'r-Rao bound (QCRB) in ordinary quantum walk (OQW) needs an entangled or parameter-dependent initial state, SSQW can outperform OQW without such requirements. As a result, for experimental and practical uses, it is possible to skip those cumbersome state preparation steps. Second, SSQW-based estimation algorithms in single-parameter estimation schemes provide a novel feature for increasing the precision through introducing an additional tunable parameter. Third, for multi-parameter estimation scenarios, SSQW surpasses OQW for a wide range of unknown parameters and variables can be estimated significantly more accurately. Furthermore, the results of this study shows that, the topological characterization of system may affect the capabilities and potency of a model for being used for estimation. Specially, we observe that the SSQW in nontrivial topological phase is mush powerful than the model in trivial topological phase. So this study can be regarded as a witness for study of the connection between the topology and quantum estimation.

\bibliography{QFISSQW}
\section{Appendix A}
By using the explicit form  of $u_k$ in Eq.\eqref{uk}, it is straightforward to show that, $A_k\left(O\right)\equiv u_{k} O u_{k}^{\dagger}$  can be written as $\tilde{A}_{k}\cket{O}$, in which
\begin{widetext}
	\begin{align} \label{A}
		\tilde{A}_{k} =  \left(\begin{array}{cccc}
			1 & 0 & 0 & 0 
			\\
			0 & C_1 C_2\cos\left(k\right)^2 - C_1\sin\left(k\right)^2 - S_1 S_2\cos\left(k\right) & -\left(1 + C_2\right)\sin\left(k\right)\cos\left(k\right) & S_1 C_2\cos\left(k\right)^2 - S_1\sin\left(k\right)^2 + C_1 S_2\cos\left(k\right)
			\\
			0 & \sin\left(k\right)\left(\left(C_1 + C_1 C_2\right)\cos\left(k\right)-S_1 S_2 \right) &  \cos\left(k\right)^2-C_2 \sin\left(k\right)^2  &  \sin\left(k\right)\left(\left(S_1 + S_1 C_2\right)\cos\left(k\right)+C_1 S_2 \right)
			\\
			0 &-S_1 C_2 -C_1 S_2\cos\left(k\right)   & S_2 \sin\left(k\right) &C_1 C_2 -S_1 S_2\cos\left(k\right)
		\end{array}\right)	 
	\end{align}
\end{widetext}
is 4-dimensional matrix and $\cket{O}$ is defined in Eq. \eqref{4vector-O} and have used $C_i=\cos\left(\theta_{i}\right)$ and $S_i=\sin\left(\theta_{i}\right)$ for simplicity.
This super-operator have four eigenvalues $\lambda_1=\lambda_2=1$ and $\lambda_3,\lambda_4=e^{\pm 2i\omega}$ with $\cos \! \left(w \right) = 
\cos \! \left(\frac{\theta_{2}}{2}\right) \cos \! \left(\frac{\theta_{1}}{2}\right) \cos \! \left(k \right)-\sin \! \left(\frac{\theta_{2}}{2}\right) \sin \! \left(\frac{\theta_{1}}{2}\right)$.

The super-operator $\tilde{A}_{k}^{\mathbb{1}}$, is the subspace of $\tilde{A}_{k}$ with eigenvalues of one, so
\begin{equation}\label{A1-lambda_i}
	\tilde{A}_{k}^{\mathbb{1}}=\cket{\lambda_1}\cbra{\lambda_1}+\cket{\lambda_2}\cbra{\lambda_2}.
\end{equation}
 By some calculation and simplification, it is not hard to find 
 \begin{equation}\label{lambda_i}
 	\cket{\lambda_i}=\frac{1}{N_i}\left(\begin{array}{c}
 		c_2 c_1 \cos\left(k \right)-s_1 s_2+\left(-1\right)^i 
 		\\
 		-c_2 s_1 \sin\left(k \right) 
 		\\
 		c_1 s_2+s_1 c_2\cos\left(k \right) 
 		\\
 		c_2 c_1\sin\left(k \right)
 	\end{array}\right),
 \end{equation}
 where $N_i=\sqrt{\left(\lambda_i|\lambda_i\right)}$ is normalization factor. By substitution of $\cket{\lambda_i}$ into Eq. \eqref{A1-lambda_i} one can find the final form of $\tilde{A}_{k}^{\mathbb{1}}$ appears in Eq. \eqref{A1}.
\end{document}